\UseRawInputEncoding
\documentclass[aps,prc,twocolumn,superscriptaddress,showpacs,showkeys]{revtex4-1}
\usepackage{graphicx}
\usepackage{amsmath}
\usepackage{lineno}
\usepackage{hyperref}

\begin{document}
%   \sloppy
%	\setpagewiselinenumbers
%	\modulolinenumbers[2]
%   \linenumbers

\title{Precision measurement of $^{\mathbf{210}}$Bi $\beta$-spectrum}

\author{I.~E.~Alekseev}
\affiliation{V.G. Khlopin Radium Institute, St.~Petersburg 194021, Russia}

\author{S.~V.~Bakhlanov}
\affiliation{Petersburg Nuclear Physics Institute, Gatchina 188350, Russia \\ National Research Center ``Kurchatov Institute''}

\author{A.~V.~Derbin}
\affiliation{Petersburg Nuclear Physics Institute, Gatchina 188350, Russia \\ National Research Center ``Kurchatov Institute''}

\author{I.~S.~Drachnev}
\affiliation{Petersburg Nuclear Physics Institute, Gatchina 188350, Russia \\ National Research Center ``Kurchatov Institute''}

\author{I.~M.~Kotina}
\affiliation{Petersburg Nuclear Physics Institute, Gatchina 188350, Russia \\ National Research Center ``Kurchatov Institute''}

\author{I.~S.~Lomskaya}
\affiliation{Petersburg Nuclear Physics Institute, Gatchina 188350, Russia \\ National Research Center ``Kurchatov Institute''}

\author{V.~N.~Muratova}
\affiliation{Petersburg Nuclear Physics Institute, Gatchina 188350, Russia \\ National Research Center ``Kurchatov Institute''}

\author{N.~V.~Niyazova}
\affiliation{Petersburg Nuclear Physics Institute, Gatchina 188350, Russia \\ National Research Center ``Kurchatov Institute''}

\author{D.~A.~Semenov}
\affiliation{Petersburg Nuclear Physics Institute, Gatchina 188350, Russia \\ National Research Center ``Kurchatov Institute''}

\author{M.~V.~Trushin}
\affiliation{Petersburg Nuclear Physics Institute, Gatchina 188350, Russia \\ National Research Center ``Kurchatov Institute''}

\author{E.~V.~Unzhakov}
\affiliation{Petersburg Nuclear Physics Institute, Gatchina 188350, Russia \\ National Research Center ``Kurchatov Institute''}

\begin{abstract}
    The precision measurement of $\beta$-spectrum shape for $^{210}$Bi (historically RaE) has been performed with a spectrometer based on semiconductor Si{(Li)} detector.
    This first forbidden non-unique transition has the transition form-factor strongly deviated from unity and knowledge of its spectrum would play an important role in low-background physics in presence of $^{210}$Pb background. 
    The measured transition form-factor could be approximated as $C(W) = 1 + (-0.4470 \pm 0.0013) W + (0.0552 \pm 0.0004) W^2$, that is in good agreement with previous studies and has significantly increased parameter precision.
\end{abstract}

%\pacs{14.60.S, 96.60.J, 26.65, 13.35.H, 13.10.+q}
%\keywords {heavy sterile neutrino, organic scintillator}

\maketitle

\section{Introduction}

Precision measurements of the $\beta$\nobreakdash-spectra are currently very important in neutron and nuclear $\beta$\nobreakdash-decay studies, as a means of searching for the effects beyond the Standard Model (SM) in the low energy region~\cite{Herczeg2001,Nico2005}.
Accurate studies of nuclear $\beta$\nobreakdash-decays have been exploited for many years in various applications of fundamental physics problems, predominantly in neutrino physics.

In this paper we present the results of the measurement of $^{210}$Bi $\beta$\nobreakdash-spectrum performed with spectrometer based on Si{(Li)}-detectors~\cite{Alexeev2018,Bazlov2018}.
The problems of $^{210}$Bi $\beta$-decay such as strong deviation from the allowed energy distribution, prolonged lifetime and anomalous  longitudinal electron polarization has has been investigated widely starting from 1930s in numerous experimental and theoretical works~\cite{Conor1937,Martin1939,Neary1940,Konopinski1941,Newby1959,Fujita1962,KIM1963,SPECTOR1963,SODEMANN1965,DAMGAARD1969,Fayans1970,Lohken1971,Morita1971,EBERT1975}.
The situation was clarified after the assumption that the ground state of $^{210}$Bi is the combination of several wave functions, calculation of nuclear matrix elements for $\beta$\nobreakdash-decay on the basis of the finite Fermi systems theory and extracting the nuclear wave functions directly from the experimental data (see~\cite{Behrens1974,GrauCarles1996} and references quoted therein).
The latest measurements of $^{210}$Bi $\beta$\nobreakdash-decay spectrum were performed in~\cite{Plassmann1954, Daniel1962, Flothmann1969, GrauCarles1996} via magnetic lens and solid state $\beta$\nobreakdash-spectrometers. The need for precise study of  $^{210}$Bi $\beta$\nobreakdash-spectrum and improved of the shape factor continues to be an important task in nuclear physics.

The bismuth isotope $^{210}$Bi belongs to the natural radioactive decay chain of $^{238}$U.
As a product of the radioactive gas $^{222}$Rn and the subsequent long-lived $^{210}$Pb, the isotope $^{210}$Bi is present inside or on the surface of almost all structural materials.
At present, the precise measurement of $^{210}$Bi $\beta$\nobreakdash-spectrum remains a crucial task for background modeling of modern neutrino detectors, as well as for the dark matter searches or other low-background experiments.
In particular, the shape of $^{210}$Bi $\beta$\nobreakdash-spectrum is very similar to the spectrum of recoil electrons originated from scattering of solar CNO-neutrinos~\cite{Agostini2019}, so in order to extract the CNO signal it is necessary to determine the shape of $\beta$\nobreakdash-spectrum with the sufficient accuracy.

\section{Experimental setup}

Magnetic~\cite{Bergkvist1972,Tretyakov1975} and electrostatic~\cite{Lobashev1985,Aker2019} $\beta$\nobreakdash-spectrometers possess the superior energy resolution, but it comes at the cost of large scale and complexity of such experimental setup.
Since the electron free path at $3$~MeV of kinetic energy (which is, basically, the maximum $\beta$\nobreakdash-transition energy for long-living isotopes) does not exceed $2$~g/cm$^3$, solid state scintillation and ionization detectors were effectively employed for detection of $\beta$\nobreakdash-electrons~\cite{Simpson1985,Derbin1993}.
The main drawback of the solid state scintillators is their relatively poor energy resolution, which stands at approximately $10\%$ at $1$~MeV as well as non-linearity effects related with quenching and emission of \v Cerenkov radiation.

In case of semiconductor detectors, there is a significant probability of back-scattering from the detector surface that depends upon the detector material.
The most widespread silicon-based semiconductors have the backscattering probability of the order of $~10\%$ for $100$~keV electrons at normal incidence~\cite{Derbin1997}.
In case of the electron energies above $1$~MeV and high $Z$ detector materials, it also becomes important to take the bremsstrahlung into account.
Still, good linearity of these detectors combined with high energy resolution gives them a lot of advantages with respect to other types of solid state detectors.

The layout of the $\beta$\nobreakdash-spectrometer used for our measurements was based on a simple ``target-detector'' geometry~\cite{Alexeev2018,Bazlov2018}.
The Si{(Li)} detector with sensitive region diameter of $15.1$~mm and thickness of $6.6$~mm was produced by standard diffusion-drift technology~\cite{Bazlov2019}.
Since the detector sensitive region did not cover the whole detector, it was fitted with a tungsten collimator ($14$~mm diameter) that ensured that incident electrons were either backscattered or stopped in the i-region of the detector.

The whole setup was located in a vacuum cryostat and cooled down to liquid nitrogen temperature.
The setup was equipped with a moderate passive shielding ($50$~mm of iron and $10$~mm of copper) that allowed for reduction of the environmental backgrounds by a factor of~7, down to $2.6\times10^{-1}$~counts/s above $50$~keV.

The detector was operated with bias voltage of $800$~V.
The readout was processed with a charge-sensitive preamplifier with resistive feedback and cooled FET transistor of the first cascade.
The preamplifier signal was processed with a standard CR\nobreakdash-3RC analogue shaper and digitized with a 14\nobreakdash-bit ADC\@.
The energy resolution determined for $59.6$~keV $\gamma$-line of $^{241}$Am turned out to be $\mathrm{FWHM} = 900$~eV for the full width at half maximum.

\begin{figure}[t]
	\centering
	\includegraphics[bb = 100 100 499 750, width=6cm,height=7cm]{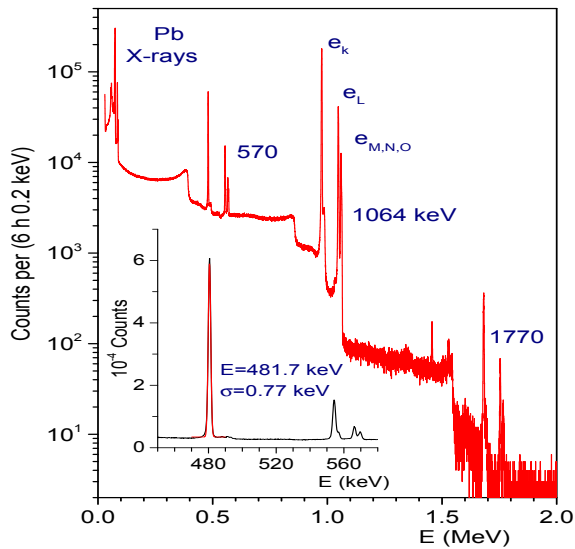}
	\caption{The spectrum of $^{207}$Bi source measured with the Si{(Li)} detector in energy range of $(0.01 - 2.0)$~MeV.
		The inset shows the electron peaks corresponding to internal conversion from $K$, $L$, $M$ and $N$ shells of the $570$~keV nuclear level.}\label{fig:Fig1}
\end{figure}

In order to determine the main characteristics of the spectrometer, we used a $^{207}$Bi source, providing $\gamma$- and X\nobreakdash-rays, conversion and Auger electrons.
The $^{207}$Bi source with an activity of $10^4$~Bq was placed inside the vacuum cryostat at a distance of $14$~mm above the Si{(Li)} detector surface.
The $^{207}$Bi spectrum, measured with the Si{(Li)} detector, is shown in Fig.~\ref{fig:Fig1} for the interval $(0.01 - 2.0)$~MeV~\cite{Alexeev2018}.

Three of the most intense $^{207}$Bi $\gamma$-lines had energies of $569.7$~keV, $1063.7$~keV and $1770.2$~keV and are emitted with probabilities of $0.977$, $0.745$ and $0.069$ per single $^{207}$Bi decay, respectively~\cite{Lederer1978,Chu1999}.
The corresponding peaks of the conversion electrons from $K$, $L$ and $M$ shells were clearly visible in the spectrum in Fig.~\ref{fig:Fig1}.
The electron energy resolution determined via $480$~keV line is $\mathrm{FWHM} = 1.8$~keV.
Energy calibration performed using Pb $K_{\alpha1}$ X-ray and $\gamma$-line with energies of $74.97$~keV and $569.70$~keV, correspondingly, predicts the position of $975.66$~keV conversion electrons peak with an accuracy better than $0.3$~keV.

The low-energy part of the $^{207}$Bi spectrum was used for evaluation of the thickness of the non-sensitive layer on the surface of Si{(Li)} detector.
This area contained a set of peaks corresponding to Pb X\nobreakdash-rays from $K$ and $L$ series and Auger electrons.
The observed position of $56.94$~keV Auger peak ($e_{K, L_{1}, L_{2}}$) appeared to be at $56.22$~keV.
Taking the $500$~{\AA} the gold coating into account, the measured energy loss of $720$~eV for $57$~keV electrons corresponded to $4700$~{\AA} of the non-sensitive layer.

The planar $^{210}\mathrm{Bi}$ source was prepared with the method of thermal oxidation~\cite{Alekseev2016}.
The polished stainless steel foil with diameter of $24$~mm  and thickness of $11$~$\mu$m was used as substrate for application of $^{210}\mathrm{Bi}$ .
A water-alcohol $^{210}\mathrm{Bi}$-containing solution was deposited onto the oxidized surface of the foil. 
The solution was air-dried and then annealed for $3$~minutes at the temperature of $300^\circ\mathrm(C)$ in order to diffuse the radioactive isotope into the oxidized surface of the substrate.

This technique is capable of producing the source of negligibly small thickness, suppressing the effects caused by the attenuation and scattering of the electrons inside the bulk material of the source itself.
The source produced in such a way decreases the systematic uncertainties of the measurement, since mentioned effects are usually difficult to simulate due to the complications with source geometry reconstruction.

\section{The results of measurements}

The natural radioactivity of the $^{238}$U and $^{232}$Th families, along with the long-lived $^{40}$K isotope, are the main sources of background for neutrino physics and dark matter searches at energies below $3 - 5$~MeV.
The main decay modes and half-life $T_{1/2}$ values of daughter nuclei produced by a long-lived  $^{210}$Pb isotope are:
\begin{equation}\label{210Pbchhain}
^{210}\mathrm{Pb}(\beta, 22.3\ \mathrm{y}) \rightarrow
^{210}\mathrm{Bi}(\beta, 5.0\ \mathrm{d}) \rightarrow
^{210}\mathrm{Po}(\alpha, 138\ \mathrm{d}).
\end{equation}
The end-point energies of the $^{210}$Pb and $^{210}$Bi $\beta$\nobreakdash-spectra are $63.5$~keV and $1162$~keV, respectively, while the energy of $^{210}$Po $\alpha$\nobreakdash-particles is $5.304$~MeV~\cite{Lederer1978,Chu1999}.
Since our $^{210}$Pb source was custom-made and intentionally purified from other lead isotopes, the equilibrium of the decay chain~(\ref{210Pbchhain}) had not yet been established at the time of measurement. 
\begin{figure}[h]
	\centering
	\includegraphics[bb = 100 100 499 750, width=6cm,height=7cm]{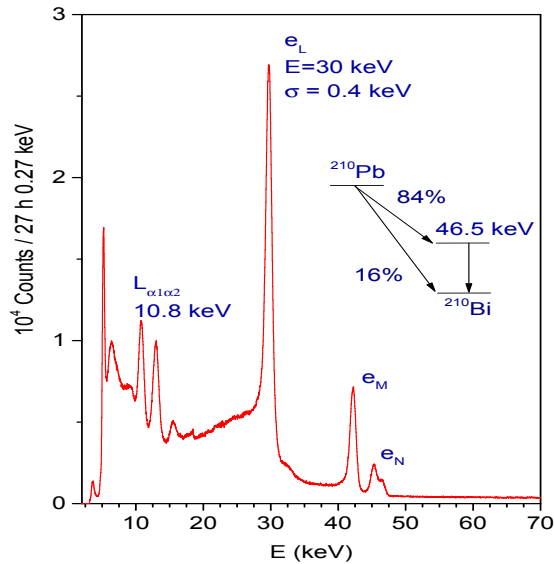}
	\caption{Low energy part of $^{210}\mathrm{Pb} \rightarrow {^{210}}\mathrm{Bi}$ spectra measured with Si{(Li)}-detector.
		The inset shows the decay scheme of $^{210}$Pb.}\label{fig:Fig2}
\end{figure}

The Fig.~\ref{fig:Fig2} shows the low-energy region of the measured spectrum determined mainly by $^{210}$Pb decays.
Transition from $46.5$~keV nuclear level of $^{210}$Bi has significant internal conversion coefficient ($e/\gamma\simeq 20$,~\cite{Lederer1978}).
Therefore, the electron peaks corresponding to conversion from $L$, $M$ and $N$ shells are clearly visible in the spectrum.

The energy resolution of Si{(Li)} detector determined for $30$~keV electron conversion line was determined to be $\mathrm{FWHM} = 1.0$~keV and lower energy detection threshold was about $5$~keV.
The kinetic energy of the recoil nucleus from $\alpha$-decay of $^{210}$Po is 100 keV. The wide peak that looks like the left shoulder of $\rm{e_L}$ peak is probably associated with these events. 
The spectrum also shows the peaks of characteristic $10.8$~keV and $13.0$~keV $L_{\alpha1}$ and $L_{\beta1}$ X\nobreakdash-rays and the wider peak of $15.5$~keV Auger $e_{LNM}$ electrons.

\begin{figure}[h]
	\centering
	\includegraphics[bb = 100 100 499 750, width=6cm,height=7cm]{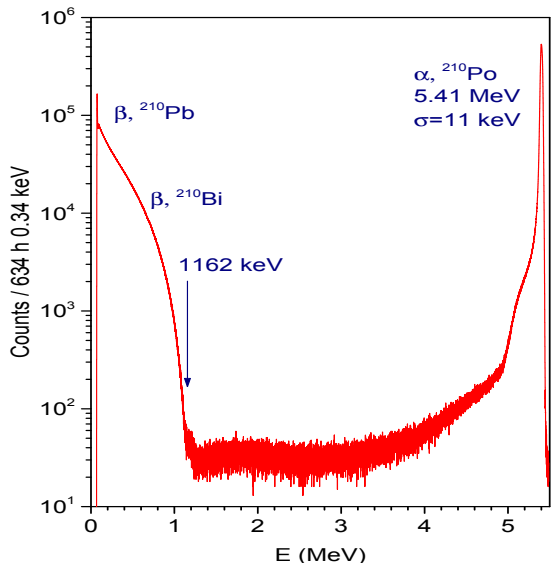}
	\caption{The energy spectrum of $^{210}\mathrm{Pb} \rightarrow {^{210}\mathrm{Bi}} \rightarrow {^{210}\mathrm{Po}}$ source measured with the Si{(Li)} detector in energy range of $(0.05-5.5)$~MeV.
		The $\beta$-spectrum of $^{210}$Bi has $1162$~keV end-point energy, the $\alpha$-decay of $^{210}$Po leads to $5.4$~MeV peak.}\label{fig:Fig3}
\end{figure}

The whole spectrum in the energy range of $(0.05-5.5)$~MeV is shown in the Fig.~\ref{fig:Fig3}.
The energy resolution of $5407$~keV $^{210}$Po $\alpha$\nobreakdash-peak was determined to be $\mathrm{FWHM} = 26$~keV. 
The peak is slightly asymmetric due to the final thickness of the target and possible other alpha impurities.
The background level near the end-point energy of $^{210}$Bi $\beta$\nobreakdash-spectrum amounted to $0.18$~counts/h/keV and that was contributed mainly by Compton scattering of $1.46$~MeV $\gamma$\nobreakdash-quanta of $^{40}$K passing through the passive shielding.
The maximum energy of recoil electrons at the edge of Compton scattering is $E_C = 2E^2/(2E+m_e)$ = 1243 keV that differs significantly from the $^{210}$Bi $\beta$\nobreakdash-decay endpoint energy.

The counting rate in the range from $80$~keV to $1.5$~MeV was 27~s$^{-1}$, that with 1~$\mu$s pile-up rejection time leads to negligible pile-up spectrum and dead time of the spectrometer.

The data was obtained during $634$~hours of data-taking in short 1-hour series used for stability control.
To determine the energy calibration $E = a + bN$ (where $E$ is a Si{(Li)} visible energy and $N$ is an ADC channel number), the position of $46.5$~keV $\gamma$-peak and the value of $^{210}$Bi end-point energy $E_0 = 1162$~keV measured with high accuracy in other experiments~\cite{Lederer1978,Chu1999} were used.

During the fitting of the spectrum, the calibration slope $b$ equal to the analyzer channel width was free, while the value of the parameter $a$ was fixed by $46.5$~keV peak position.
The differences of fitting parameters for the all 1-hour runs are in agreement with their statistical uncertainties.
The fact that equilibrium in (\ref{210Pbchhain}) was not achieved could not affect the fitting results for different series, if only because the contribution of the tail of $\alpha$\nobreakdash-particles to the $\beta$\nobreakdash-spectrum region was very small.
The total number of registered $^{210}$Bi decays was $1.0\times10^8$.

\section{Data analysis}

The energy distribution $S(W)$ of $\beta$-particles emitted in $\beta$-decay process could be expressed as
\begin{equation}
S(W) = P W {(W - W_0)}^2 \times F(W, Z) \times C(W),
\end{equation}
where $P$ and $T$ are the electron momentum and energy, $W = T/mc^2 + 1$ is full electron energy, $W_0 = T_0 / mc^2 + 1$ is $\beta$-spectrum end-point energy, $F(W, Z)$ is the electron Fermi function that takes into account electromagnetic interaction of the outgoing electron with the atom and $C(W)$ is the transition nuclear form-factor that considers the effects of internal nuclear interactions.

The Fermi function $F(W,Z)$ is historically derived in approximation of a point-like nuclei without consideration of the atomic shells~\cite{Fermi1934} that means that comparison with experiments using this model needs application of the same approximation, while the $\beta$-spectrum for practical applications would need a more profound calculation of the Fermi function that was performed according to~\cite{Dzhelepov1956,Huber2011,Hayen2018}.

The transition investigated in this work is of forbidden type and the nuclear form-factor $C(W)$ is expected to deviate from unity and is the main subject of the measurement.
Since the shape factor of first forbidden non-unique transition with such parity-momentum relations can be expressed with sufficient accuracy by a second degree polynomial, we choose the $C(W)$ parametrization as in~\cite{GrauCarles1996}:
\begin{equation}\label{eq:FFactor}
C(W) = 1 + C_1 W + C_2 W^2
\end{equation}
with generic values of parameters $C_1$ and $C_2$ that were defined through maximum likelihood fit with $\chi^2$ likelihood function.

The final model of the experimental spectrum expresses as:
\begin{equation}
N(E) = \int_{E/mc^2 + 1}^{W_0} S(W)\times R(W,E)dW,
\end{equation}
where $R(W, E)$ is the spectrometer normalized response function obtained with Monte-Carlo simulation of electrons with energy $W$ exiting the source with uniform distribution within the source and uniform distribution of their momenta directions.

Since the setup in use has the classical ``target-detector'' geometry, it is quite important to take into account the detector response function that would contain a long low-energy tail caused by fraction of electrons backscattered from the detector as well as by bremsstrahlung exit from the detector crystal.
The Si{(Li)} detector has i-region thickness exceeding the stopping range of an electron with endpoint energy of $1162$~keV and thus the geometry of irradiated regions of the setup is quite well established.
This allows to account for the detector energy response through a precise simulation with the Monte-Carlo method.
We used Geant4.10.04 simulation package~\cite{GEANT4} with the standard G4EmStandardPhysics\_option4  package of electromagnetic interactions.

The package choice was mainly motivated by the Single Scattering model for electrons, that is the most promising among standard ones according to~\cite{Basaglia2016, Dondero2018}.
The simulation was including modeling of the detector entrance windows, collimator and holders according to the physical setup geometry.

As the response function model is based on the simplified interaction models used in the simulation, it is important to estimate the uncertainties concerning its imperfection.
Consideration of these uncertainties was performed through analytical modification of the response function as:
\begin{equation}
\tilde{R}(E,W) =
\begin{cases}
R(E,W)\times(1 + A\,\ln(B\,W)),&  \\ \ \ \ \ \ \ \text{if } E < T - 5\sigma \\
R(E,W),&   \\ \ \ \ \ \ \ \text{if } E > T - 5\sigma
\end{cases}
\end{equation}
where $\sigma$ is the detector resolution at kinetic energy $T$ and $A$,$B$ are free parameters.
Eventually, six parameters were free in the fit: the common normalization coefficient, the slope of the energy calibration, the form factor parameters $C_1$ and $C_2$, and response function parameters $A$ and $B$.

The dependence $A \,\ln(B\,W)$ used for the variation of the response function tail approximately corresponds to the uncertainties of the response function for different GEANT4 simulation packages~\cite{Basaglia2016}.
The response function was renormalized to conserve detection efficiency of the original simulation.

The fit range has the lower bound that comes from presence of $^{210}$Pb in the source that covers the low-energy region.
Considering that the nuclear form-factor $C(W)$ depends only upon momenta of the electron and neutrino one should not expect sudden behavior in the lower tenth of the energy spectrum so this lower bound should not be important for the form-factor establishment.

The fit with canonical Fermi function $F_0(W, Z)$ was performed in the energy range $120 - 1175$~keV with flat background approximation.
The Fermi function was calculate according to~\cite{Fermi1934} as:
\begin{equation}\label{FF0}
F_0(W, Z) =
4 {(2pR(A))}^{2(\gamma - 1)} e^{(\pi Y)}
\frac{|\Gamma(\gamma + iY)|}{\Gamma^2(1+2\gamma)}.
\end{equation}
were $Y=\alpha ZW/p$ and $\gamma=\sqrt{1-\alpha^2Z^2}$, $\alpha$ is the fine-structure constant and $R$ is the nuclear radius defined as $R = 0.0029\times A^{1/3} + 0.0063\times A^{2/3} - 0.017\times A^{-1}$.

\begin{figure}[t]
	\centering
	\includegraphics[bb=0 0 520 600,width=\linewidth]{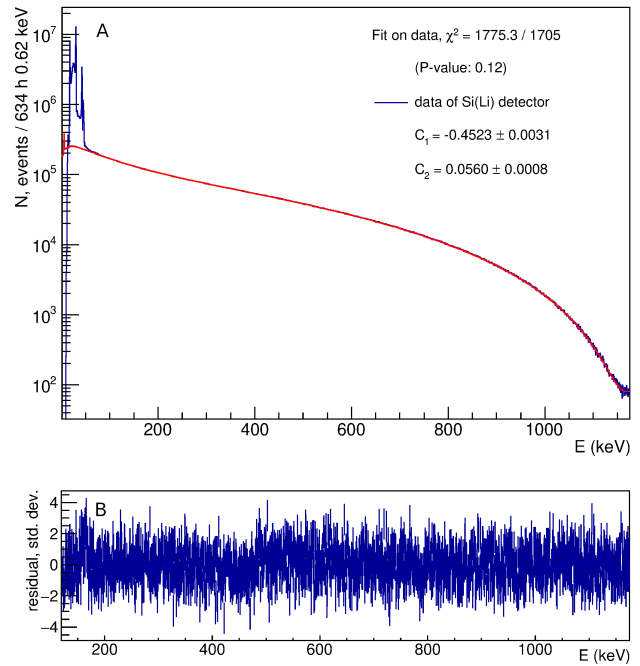}
	\caption{Experimental spectrum fit with parabolic form-factor $C(W)$ and Fermi function $F_0(W,Z)$, computed in approximation of a point-like nucleus~\cite{Dzhelepov1956}.
		The $\chi^2$ fit was performed in the energy range $120 - 1175$~keV with flat background approximation.}\label{fig:Fig4}
\end{figure}

The fit results  are shown on fig.~\ref{fig:Fig4}.
The obtained minimum of $\chi^2 / \mathrm{NDF} = 1775.3 / 1705$  corresponds to  Pearson $\mbox{P-value} = 0.12$ and form-factor parameters $C_1 = -0.4523 \pm 0.0031$ and $C_2 = 0.0560 \pm 0.0008$.
The easily computed values of $F_0(W, Z)$ (\ref{FF0}) together with  obtained coefficients $C_1$ and $C_2$ allows to calculate the shape of the $\beta$-spectrum that references to our measurements.

In order to have a fair comparison with results of~\cite{Daniel1962,GrauCarles1996} we performed the fit with the Fermi function $F(W, Z)$ calculated in accordance with formalism presented in~\cite{Dzhelepov1956}, attempting to improve the precision of the analytical description.
In this work $F(E, Z)$ was enhanced by including second and third terms of $pr$-power expansion of electron wave function at small values of $r$ ($F_0(E, Z)$ is obtained by neglecting all but the first term).
Also, additional corrections were included, taking the finite size of the nucleus and atomic shell screening into account~\cite{Sliv1947,Rose1936}.
The values of $F(E, Z) = F_0(E, Z) \cdot \chi \cdot \eta$ used in the calculation were taken from the Table 14 of~\cite{Dzhelepov1956} for $Z = 83$ and $A = 210$.

\begin{figure}[t]
	\centering
	\includegraphics[bb=0 0 520 600,width=\linewidth]{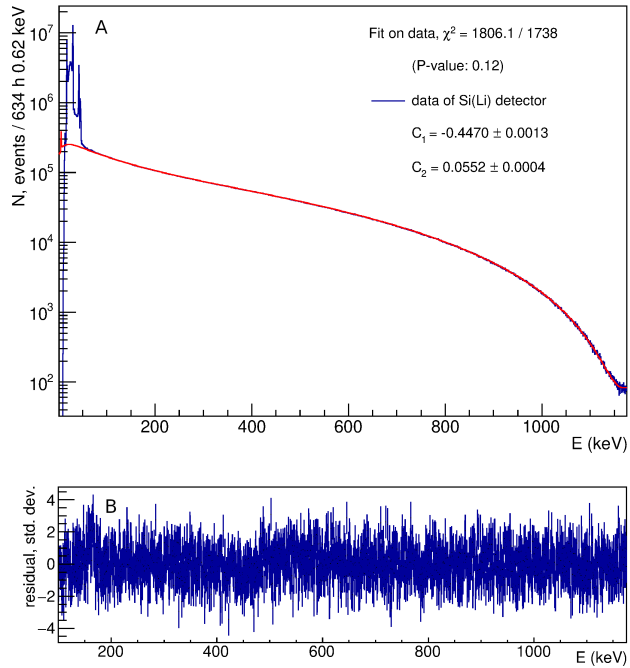}
	\caption{Experimental spectrum fit with parabolic form-factor $C(W)$ and Fermi function $F(W, Z)$, computed according to~\cite{Huber2011,Buhring1984}.
		The $\chi^2$ fit was performed in the energy range $100 - 1175$~keV with flat background approximation.}\label{fig:Fig5}
\end{figure}

The fit range was increased with respect to the improved $F(W,Z)$ that takes into account the nucleus final size and the screening corrections.
The same procedure gives the form-factor parameters as $C_1= -0.4339 \pm 0.0012$ and $C_2 = 0.0513 \pm 0.0004$.
These values one can compare with $C_1' = -0.46 \pm 0.01$ and $C_2'= 0.0586 \pm 0.002$ obtained in~\cite{GrauCarles1996}.
The errors of $C_1$, $C_2$ obtained in the present work are more than five times less, however, the parameters $C_1$, $C_2$ and $C_1'$, $C_2'$ are consistent with each other within the $1.5$~$\sigma$.

The final fitting procedure was repeated using the classic definition of the Fermi function with several corrections that included atomic shell screening effect~\cite{Buhring1984}, finite size distribution of electromagnetic and weak charge inside nucleus~\cite{Wilkinson1990} and QED radiative corrections~\cite{Sirlin1967,Hayen2018}.
The final $F(E, Z)$ had the following form:
\begin{align}
F(E, Z)=& F_0(E, Z) \times S(E, Z) \times\\
&L_0(E, Z) \times M(E, Z) \times G_{\beta}(E)\nonumber
\end{align}
where $E$ is full electron energy, $Z$ is the charge of a daughter nucleus, $F_0(E, Z)$ is Fermi function, $S(E, Z)$ screening correction, $L_0(E, Z)$ and $M(E, Z)$ are electromagnetic and weak finite size corrections and $G_{\beta}(E)$ is radiative correction.
The results of the final fitting procedure are given in Fig.~\ref{fig:Fig5}

Implication of a more precise Fermi function allowed to lower energy threshold down to 100 keV, providing good P-value = 0.12 that is an evidence of better agreement of the corrected beta-spectrum with the experimental data.
The minimum of $\chi^2/NDF$ = 1806.1 / 1738  corresponds to form-factor parameters $C_1= -0.4470 \pm 0.0013$ and $C_2 = 0.0552 \pm 0.0004$.
These values $C_1$ and $C_2$ are obtained taking into account the most complete knowledge of the interactions emitted electron with atom.
One should note that the parameters $C_1$ and $C_2$ have quite strong correlation in the fit of the experimental data, having the correlation coefficients of 0.987 (Fig.~\ref{fig:Fig4}) and 0.96 (Fig.~\ref{fig:Fig5}). The inclusion of quadratic term in the energy calibration and an additional pull term in the fitting procedure leads to increase of C1 and C2 errors by less than $5\%$ without changing the central values.

\begin{figure}[t]
	\includegraphics[bb=20 0 510 590,width=1\linewidth]{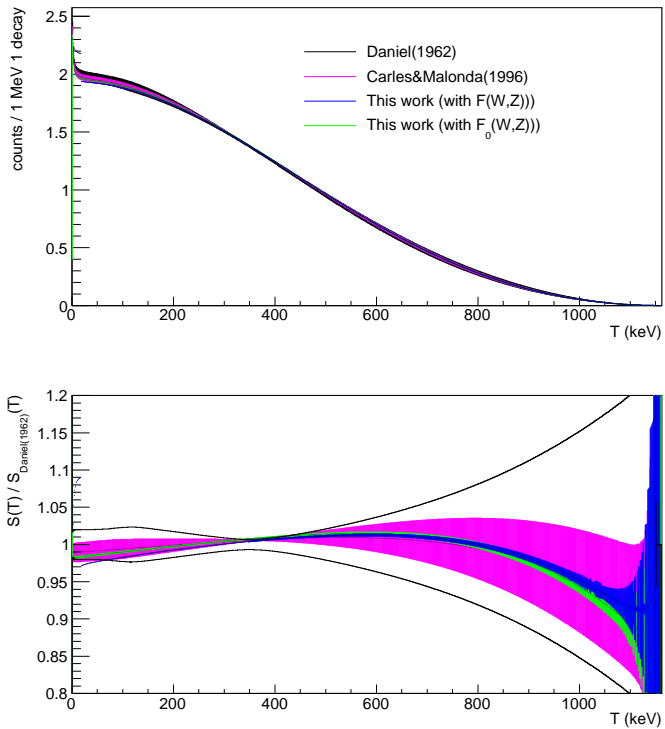}
	\caption{Comparison of the spectra measured by Daniel (1962)~\cite{Daniel1962} and Carles \& Malonda (1996)~\cite{GrauCarles1996} with the present study (top).
		Ratio to Daniel (1962) spectrum (bottom).
		Daniel (1962) spectrum errors are shown by solid black lines.}\label{fig:Fig6}
\end{figure}

The fits performed have shown convergence of parameter $A$ to null, showing that with the current experimental statistics the Monte-Carlo simulation of the spectrometer response gives results consistent with the data.
The conservative approach demonstrates the fact that inclusion of parameters A and B in the fit leads to increase errors of $C_1$ and $C_2$ by approximately a factor of three.

The present results are compared with obtained in works~\cite{Daniel1962} (Daniel (1962)) and~\cite{GrauCarles1996} (Carles \& Malonda (1996)) in Fig.~\ref{fig:Fig6}.
The upper part of the figure shows the electron spectra.
To determine the Daniel (1962) spectrum, we used the data from Table~2~\cite{Daniel1962} and the Fermi function from~\cite{Dzhelepov1956}.
The Carles \& Malonda (1996) spectrum was calculated in accordance with the parameters $C_1$ and $C_2$ given in~\cite{GrauCarles1996}.
The figure shows two electron spectra obtained in the present work for $F_0(W,Z)$ and $F(W,Z)$ Fermi functions according to~\cite{Fermi1934} and~\cite{Huber2011,Buhring1984}.
All $\beta$-spectra were normalized to unity.

Fig.~\ref{fig:Fig6} shows also the ratio of Carles \& Malonda (1996) and present work spectra to Daniel (1962) spectrum~\cite{Daniel1962}.
Daniel (1962) spectrum errors determined by fit are shown by solid black lines.

Because $C_1$ and $C_2$ have quite strong correlation, in order to estimate uncertainties on the form factor curve shown at Fig.~\ref{fig:Fig6} we applied the Monte-Carlo method sampling the form-factor parameters according to two-dimensional Gaussian distribution that includes the correlation coefficient obtained in the fit.
One can see that both of our spectra are consistent with Daniel (1962) and Carles \& Malonda (1996) spectra within uncertainties.
The current study shows significantly increased precision with respect to the previous studies.

\section{Conclusions}

The spectrometer based on the Si{(Li)} detector was used to precisely measure the $\beta$-spectrum of $^{210}$Bi nuclei.
As a result of the 634 hours measurements with a total number of $1.0\times10^8$ of registered electrons it was established that the $\beta$-spectrum is described by form-factor $C(W) = 1 + (-0.4523 \pm 0.0031) W + (0.0560 \pm 0.0008) W^2 $ if the Fermi function is calculated according to formula~(\ref{FF0}) for a point-like nucleus.
The obtained values of the parameters $C_1$ and $C_2$ together with~(\ref{FF0}) can be used for calculation of the electron spectrum of $^{210}$Bi.

When the additional above-mentioned corrections to the Fermi function are taken into account, the form-factor parameters are equal  $C_1=(-0.4470 \pm 0.0013)$  and  $C_2=(0.0552 \pm 0.0004)$, that can be useful for calculation of specific nuclear matrix elements.
The obtained parameters of the form-factor are in agreement with the previous studies and have significantly increased  precision.

\section{Acknowledgments}
This work was supported by the Russian Foundation for Basic Research (project nos. 19-02-00097 and 20-02-00571).

%\bibliography{ref_fix}
%\bibliography{210Bi}
%\bibliographystyle{unsrtnat}

\end{document}